\documentclass[fleqn,twoside]{article}
\usepackage{espcrc2,latexsym,amssymb,mathbbol}

\usepackage{graphicx}

\newcommand{\AmS}{{\protect\the\textfont2
  A\kern-.1667em\lower.5ex\hbox{M}\kern-.125emS}}

\newcommand{\be}{\begin{equation}}
\newcommand{\ee}{\end{equation}}
\newcommand{\bea}{\begin{eqnarray}}
\newcommand{\eea}{\end{eqnarray}}
\newcommand{\bml}{\begin{mathletters} \baselineskip 10pt}
\newcommand{\eml}{\baselineskip 12pt \end{mathletters}}

\newcommand{\mrm}{\mathrm}

\newcommand{\la}{\lambda}

\newcommand{\de}{\delta}

\newcommand{\bra}{\langle}
\newcommand{\ket}{\rangle}

\newcommand{\simgeq}{\scriptstyle{\stackrel{>}{\sim}}}

\newcommand{\pa}{\partial}

\newcommand{\vc}[1]{\mbox{\bf #1}}

\newcommand{\tr}{\mbox{tr}}

\title{Effective theories of confinement\thanks{
Talk presented by T.H.
}
}

\author{L.~Dittmann\address[TPI]{Theoretisch--Physikalisches Institut, 
        Friedrich--Schiller--Universit\"at Jena, Max--Wien--Platz 1,
        07743 Jena, Germany},
        T.~Heinzl\addressmark\address{Sektion Physik,
        Ludwig--Maximilians--Universit\"at, Theresienstra\ss e 37, 80333
        M\"unchen,
        Germany}\thanks{heinzl@theorie.physik.uni-muenchen.de}\thanks{supported in part by DFG},  
        A.~Wipf\addressmark[TPI]}

\begin{document}

\begin{abstract}
We review some approaches to describe confinement in terms of
effective (model) field theories. After a brief discussion of the dual
Abelian Higgs model, we concentrate on a lattice analysis of the
Faddeev--Niemi effective action conjectured to describe the low--lying 
excitations of $SU(2)$ gluodynamics. We generalize the effective
action such that it contains all operators built from a unit color
vector field $n$ with $O(3)$ symmetry and maximally four
derivatives. To avoid the presence of Goldstone bosons, we include
explicit symmetry breaking terms parametrized by an external field $h$
of mass--dimension two. We find a mass gap of the order of 1.5 GeV.

\end{abstract}


\maketitle

\section{INTRODUCTION}

Let us begin by recalling the definition of confinement: in
\textit{pure} Yang--Mills theory, the potential between static quarks
in the fundamental representation ($t = 1/2$ for $SU(2)$) grows
\textit{linearly} at large interquark distances, $R \; \simgeq$ 1 fm,
\be 
  V_{Q \bar Q} = \sigma_{1/2} R \; .
\ee
The constant of proportionality is the (fundamental) string tension, 
$\sigma_{1/2} \simeq \, $(450 MeV)$^2$. The definition above is
equivalent to the statement that the Wilson loop decays with an area
law. 

For static sources in higher representations ($t = 1, 3/2, \ldots$)
one has \textit{screening} for large $R$. For intermediate distances,
however, there is again a linear rise, $V_t = \sigma_t R$, where the
higher string tensions obey `Casimir scaling'
\cite{feynman:76,ambjorn:84,bali:99}, 
\be
  \sigma_t = C_t \sigma \equiv t(t+1) \sigma \; , 
\ee
A potential $V_t$ proportional to the quadratic Casimir $C_t$ is, of
course, well known from perturbation theory. The one--gluon exchange
potential is $V_\mathrm{OGE} \sim $ (charge)$^2 \sim T^a T^a$. It is,
however, unclear why such a behavior should also be true at larger
distances, i.e.~in the nonperturbative regime.

The simplest model explanation of confinement is the
dual--superconductor scenario \cite{thooft:76a,mandelstam:76}. This
views the QCD vacuum as a monopole condensate, for which a dual
Meissner effect leads to a (chromo)electric flux tube between static
quarks. 

Apart from (static) quark confinement there should also be gluon
confinement implying a finite range of the gluonic interactions,
i.e.~a mass gap. The connection between the linear potential and the
existence of a mass gap is somewhat elusive. A qualitative Peierls type
argument goes as follows \cite{polyakov:87}. The partition function
for magnetic flux lines of length $L$ has an energy--entropy behavior
according to
\be
\label{Z}
  Z \sim \exp(-cL/e^2 + L \, \ln c') \; , 
\ee
where we have used electric--magnetic duality in writing the magnetic
coupling as $1/e$. Energy--entropy balance suggests a phase transition
(monopole condensation) if $e \sim 1$ leading to `topological order'
\cite{kosterlitz:73}. This implies \textit{disorder} of the dual objects which, 
roughly speaking, should be the `glue'. Thus, there is no long--range
order in the gluonic sector so that there must be a mass gap.

\section{EXAMPLES}

In this section we will discuss two alternative approaches,
emphasizing either linear confinement or the presence of a mass gap.

\subsection{Dual Abelian Higgs model (DAHM)}

This model \cite{baker:85,maedan:89} is basically a field theoretical
Ginzburg--Landau realization of the dual superconductor. Thus, the
Lagrangian is
\be
  \mathcal{L} = - \frac{1}{4g^2}F^2 + |D\phi|^2 + \lambda (|\phi|^2 -
  \phi_0^2)^2 \; .
\ee
the ingredients being a dual photon coupled to a magnetically charged
Higgs field, $\phi$. Dual superconductivity is achieved by the familiar
Higgs mechanism. The Higgs vacuum expectation value, $\bra \phi \ket
\equiv \phi_0$, provides the scale for both the photon and Higgs mass,
$m_\gamma^2 = 2 g^2 \phi_0^2$, $ m_H^2 = 4 \lambda \phi_0^2$.  Note
that with $\phi_0^2$ there is a new operator of mass--dimension two in
the game \cite{shuryak:99}.

The model has a classical soliton solution, the
Abrikosov--Nielsen--Olesen (ANO) vortex, which has finite energy per
unit length implying a linear potential, $V(L) = \sigma L$. The string
tension $\sigma$ is proportional to $\phi_0^2$.

Interestingly, it is possible to derive a string representation of the
ANO vortex \cite{foerster:74,gervais:75,orland:94}, for a review see
\cite{antonov:00}. One obtains a Nambu--Goto action with higher
curvature terms added. 

Like any model, the DAHM also has its problems. Being Abelian, it
actually describes a $U(1)$ confinement. This can be viewed as
stemming from an Abelian projection of Yang--Mills theory
\cite{thooft}, where the $SU(2)$ gauge freedom has been partially
fixed down to a $U(1)$ subgroup. For this reason there is no
confinement of objects that are uncharged (neutral) with respect to
this $U(1)$, for example the diagonal gluons. In addition, it turns
out that there is no Casimir scaling in this model. This has been
rectified only recently \cite{chernodub:01}. Finally, it seems
difficult to describe the gluonic sector in the DAHM. Attempts to
represent glueballs as closed vortices are still in their infancy
\cite{koma:99}. Thus, it is unclear how to obtain the mass gap in the
glueball spectrum. The second model we are going to discuss is meant
to do better in this particular respect.

\subsection{Faddeev--Niemi action}

Recently, Faddeev and Niemi (FN) have suggested that the
infrared sector of Yang--Mills theory might be described by the
following low--energy effective action \cite{faddeev:99a},
\be
\label{FN_ACTION}
  S_{\mrm{FN}} \!  =  \!\!\! \int \!\! d^4 x \! \left[ m^2
  (\partial_\mu \vc{n})^2\! +\!  \frac{1}{4e^2} (\vc{n} \cdot
  \partial_\mu \vc{n} \times \partial_\nu \vc{n})^2 \right]\!\! .
\ee
Here, $\vc{n}$ is a unit vector field with values on $S^2$, $\vc{n}^2
\equiv n^a n^a = 1$, $ a = 1,2,3$; $m^2$ is a dimensionful and $e$  a
dimensionless coupling constant.  The FN `field strength' is defined as
\be
  H_{\mu\nu}  \equiv \vc{n} \cdot \partial_\mu \vc{n}\times
  \partial_\nu \vc{n} \; .
\ee
FN claim that (\ref{FN_ACTION}) ``is the \textit{unique} local and
Lorentz--invariant action for the unit vector $\vc{n}$ which is at
most quadratic in time derivatives so that it admits a Hamiltonian
interpretation and involves \textit{all} such terms that are either
relevant or marginal in the infrared limit''.

It has been shown that $S_{\mrm{FN}}$ supports string--like knot
solitons \cite{faddeev:97,battye:98}, characterized by
a topological charge which equals the Hopf index of the map
$\vc{n}:S^3 \to S^2$.  In analogy with the Skyrme model,
the $H^2$ term is needed for stabilization.  The knot solitons can
possibly be identified with gluonic flux tubes and are thus
conjectured to correspond to glueballs.  For a rewriting in terms of
curvature free $SU(2)$ gauge fields and the corresponding
reinterpretation of $S_{{\rm FN}}$ we refer to \cite{Wipf}.

In order for the model to really make sense, however, the following
problems have to be solved.  First of all, neither the interpretation of
$\vc{n}$ nor its relation to Yang--Mills theory have been
clarified. An analytic derivation of the FN action requires an
appropriate change of variables, $A \to (\vc{n}, X )$, which
decomposes the Yang--Mills potential $A$ into (a function of) $\vc{n}$
and some remainder $X$. Although progress in this direction has been
made
\cite{langmann:99,shabanov:99a,shabanov:99b,gies:01}, there are no
conclusive results up to now.

Second, there is no reason why both operators in the FN `Skyrme
term', which can be rewritten as
\be
H^2 = (\pa_{\mu}\vc{n})^4 + (\pa_{\mu}\vc{n} \cdot \pa_{\nu}\vc{n})^2 \; , 
\ee
should have the same coupling. Third, and conceptually most important,
$S_{\mrm{FN}}$ has the same spontaneous symmetry breaking pattern as
the non-linear $\sigma$-model, $SU(2)\to U(1)$. Hence, it should admit
two Goldstone bosons and one expects to find \textit{no} mass gap. 

We have scrutinized the FN action using lattice methods. To this end
we made a sufficiently general ansatz for an $\vc{n}$--field action
that contains (\ref{FN_ACTION}) as a special case. In particular, we
allow for explicit symmetry breaking terms to avoid the appearance of
Goldstone bosons. 

\section{LATTICE TEST}

After generating $SU(2)$ lattice configurations using the standard
Wilson action we fix to a covariant gauge following the continuum
approach of \cite{shabanov:99b,gies:01}.  We chose Landau gauge
(LG), defined by maximizing $\sum_{x,\mu}\tr\,^\Omega U_{x, \mu}$
w.r.t.~the gauge transformation $\Omega$, leaving a residual global
$SU(2)$--symmetry. The field $\vc{n}$ is then obtained via maximizing
the functional $F_\mathrm{MAG} \equiv \sum_{x,\mu} \tr \left( \tau_3
\, ^gU_{x,\mu} \tau_3 \, ^gU_{x,\mu}^\dagger \right)$ of the maximally Abelian 
gauge (MAG) \cite{thooft,wiese}. This yields a gauge transformation
$g$ which we use to define our $\vc{n}$--field,
\bea
\label{NDEF}
n_x = g^\dag_x \, \tau_3 \, g_x\ .
\eea
It is important to note that this definition leaves a residual local $U(1)$
unfixed. 

\begin{figure}[ht]
  \hspace{0.15\textwidth}
  \includegraphics[height=6cm,width=0.2\textwidth]{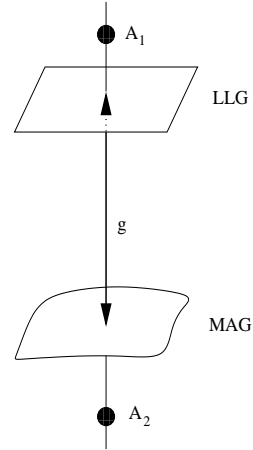}
  \caption{Gauge invariant definition of $n \equiv g^\dagger
  \tau_3 g$. The gauge equivalent configurations $A_1$ and $A_2$ are
  both mapped onto the same `representatives' on the lattice LG or MAG 
  slices (ignoring Gribov copies). Thus, they are both associated with
  the \textit{same} gauge transformation $g$ defining $n$.}
  \label{Fig.LLG_MAG}
\end{figure}

Since the configurations generated originally are randomly distributed
along their orbits, the gauge fixing is absolutely crucial for
rendering the definition (\ref{NDEF}) gauge invariant \cite{deforcrand}. This is illustrated in Fig.~\ref{Fig.LLG_MAG}.

Our ansatz for the effective action is
$S_{\rm{eff}}=\sum_i\la_iS_i[\vc{n}]$ with couplings $\la_i$ and
operators $S_i$. Up to fourth order in a gradient expansion there are
the symmetric terms
\bea
\label{SYMM}
(\nabla_{\mu}\vc{n})^2 \; ,\ (\square\vc{n})^2 \; ,\ (\nabla_{\mu}\vc{n})^4
\; ,\ (\nabla_{\mu}\vc{n} \cdot \nabla_{\nu}\vc{n})^2\ ,
\eea
and the symmetry  breaking terms including a `source field' $\vc{h}$,
\bea
\label{NONSYMM}
\vc{n}\cdot\vc{h} \; ,\ (\vc{n}\cdot\vc{h})^2 \; ,\
(\nabla_{\mu}\vc{n})^2\vc{n}\cdot\vc{h} \ .
\eea

The couplings $\la_i$ can be obtained by use of an inverse Monte Carlo
method \cite{Parisi}: rotational invariance of the functional measure
implies an infinite set of Schwinger--Dyson equations providing an
overdetermined linear system for the couplings,
\be
\label{IMC}
  \sum_j \bra F_i^{ab}[\vc{n}] S^j_{,b}[\vc{n},\vc{h}] \ket \la_j = \bra
  I^a_i[\vc{n}] \ket \; .
\ee
Here, $F_i^{ab}$ and $I^a_i$ are known functions of $\vc{n}$, typically
linear combinations of n-point functions.

\begin{figure}[ht]
  \includegraphics[width=0.469\textwidth]{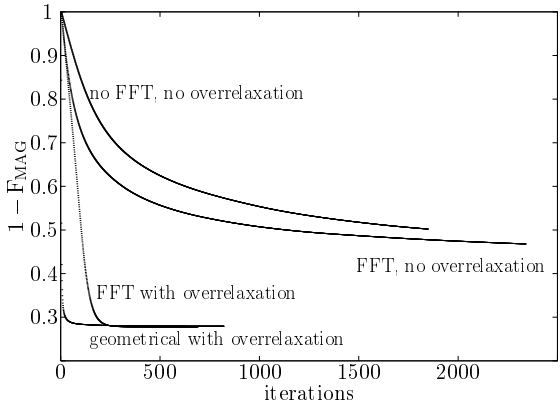}
  \vspace*{-10mm}
  \caption{Behavior of the MAG--functional using
  different algorithms.}
  \label{Fig.MAGS}
\end{figure}

All computations have been done on a $16^4$--lattice with Wilson coupling
$\beta=2.35$, lattice spacing $0.13\ {\rm fm}$ and periodic boundary
conditions.  For the LG we used Fourier accelerated steepest descent
\cite{Davis}. The MAG was achieved using two independent
algorithms, one (AI) being based on 'geometrical' iteration \cite{Bali:privat},
the other (AII) analogous to LG fixing (see Fig.~\ref{Fig.MAGS}).

\begin{figure}[ht]
  \includegraphics[width=0.469\textwidth]{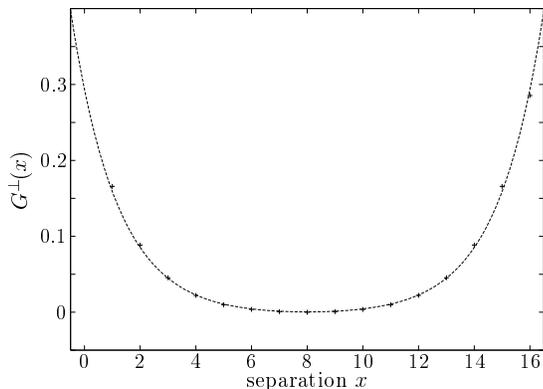}
  \vspace*{-15mm}
  \caption{The transverse correlation function
  $G^\perp$, fitted to $G^\perp (x) \sim \cosh(-M(x-L/2))$.}
  \label{Fig.coshfitt}
\end{figure}

As expected, we observe a non-vanishing expectation value of the field
in the three-direction that can be thought of as a 'magnetization'
$\mathfrak{M}$, $ \bra n^a \ket = \mathfrak{M}
\, \de^{a3}$. Thus, the global symmetry is broken explicitely according to the
pattern $SU(2)\to U(1)$. This also shows up in the behavior of the
two--point functions (Fig.~\ref{Fig.twopointfunctions}), which exhibit
clustering, $\bra n^3(0) n^3(x) \ket \sim  \bra n^3 \ket \bra n^3 \ket =
\mathfrak{M}^2$, for large distances. Furthermore, the transverse
correlation function (of the would-be Goldstone bosons)
\vspace*{-1mm}
\bea
G^\perp(x) \equiv \frac{1}{2} \bra n^i(0)n^i(x) \ket,\ i=1,2\ ,
\eea
decays exponentially as shown in Fig.~\ref{Fig.coshfitt}. This means
that there is a nonvanishing mass gap $M$ whose value can be obtained
by a fit to a $\cosh$--function. 

The numerical values of the observables, $\mathfrak{M}$, $M$ and the
transverse susceptibility, $\chi^\perp \equiv \sum_x G^\perp(x)$, are
summarized in Table 1 for both algorithms:
\begin{figure}[ht]
  \includegraphics[width=0.469\textwidth]{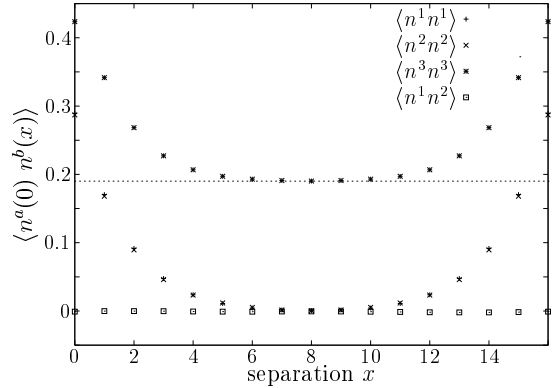}
  \vspace*{-15mm} 
  \caption{Two-point correlators of the field $\vc{n}$
  obtained via algorithm AI. The dotted line represents the (squared)
  VEV of $\vc{n}$, $\bra n^3 \ket^2 = \mathfrak{M}^2.$ The same behavior
  is obtained via AII with slightly different plateau value (see
  Table~1).}
  \label{Fig.twopointfunctions}
\end{figure}
\begin{table}[ht]
\caption{Numerical values for some observables (all numbers in units
of the lattice spacing).}
\label{table:1}
\newcommand{\hp}{\hphantom{$-$}}
\newcommand{\cc}[1]{\multicolumn{1}{c}{#1}}
\renewcommand{\tabcolsep}{0.99pc} 
\renewcommand{\arraystretch}{1.2} 
\begin{tabular}{@{}llll}
\hline
   & {$\mathfrak{M}$} & {$\chi^\perp$} & $M$  \\
\hline
AI & 0.436 & 0.636 & 0.95 \\  
AII& 0.352 & 0.596 & 1.01 \\ \hline
\end{tabular}
\end{table}
The slight disagreement between AI and AII is expected from our
still somewhat low statistics. The numerical results for the mass gap
$M$ lead to a value of about 1.5 GeV in physical units.  

A first check of our method is to consider the minimal ansatz
consisting of the first (leading) terms of (\ref{SYMM}) and
(\ref{NONSYMM}), respectively,
\be
\label{MIN_ANS}
  S_{\mathrm{eff}} = \sum_x \, \left(\la_1 (\nabla_\mu \vc{n}_x)^2 +
  \la_2 \, \vc{n}_x \cdot \vc{h}/|\vc{h}|  \right) \; , 
\ee
where $\la_2 \equiv \la_1 |\vc{h}|$.  We thus have two couplings,
$\la_1$ and $|\vc{h}|$, the latter representing an alternative new
operator of mass--dimension two. Within the ansatz (\ref{MIN_ANS}), it
can be determined using an exact lattice Ward identity and the data
from Table~1,
\be
  |\vc{h}| = \mathfrak{M}/\chi^\perp \simeq (1.3 \; \mbox{GeV})^2 \; .
\ee
In terms of the mass gap, on the other hand, one has
\be
  |\vc{h}| = M^2 + O(\la_1^{-1}) = (1.5 \; \mbox{GeV})^2 +
  O(\la_1^{-1}) \, , 
\ee
so that we find qualitative agreement already to leading order in the
derivative expansion. 

The effective couplings have to be determined by solving
(\ref{IMC}). Results already obtained will be reported elsewhere. 

\section{CONCLUSIONS}

We have discussed two effective theories meant to describe the
confinement dynamics of pure Yang--Mills theory, the dual Abelian
Higgs and the Faddeev--Niemi model. The latter has been modified
by allowing for symmetry breaking terms in order to avoid the
appearance of Goldstone modes. Both models then contain new operators
of mass--dimension two, the vacuum expectation value of the Higgs
field (squared), or an external `source' field $h$, respectively. The
former implies a non--vanishing string tension, the latter a  mass
gap.  

At the moment, the relation between the two alternative descriptions
is unclear. One may speculate that some kind of duality could unify
them.

\section*{Acknowledgements}

The authors are indebted to S.~Shabanov for suggesting this
investigation and to P.~van Baal for raising the issue of Goldstone
bosons. T.H. thanks A.~Bassetto and his team for organizing such a
stimulating meeting.  Discussions with P.~van Baal, P.~de~Forcrand,
E.~Seiler and M.~Teper are gratefully acknowledged.

\end{document}